\documentclass[12pt,english]{article}
\usepackage{lmodern}
\usepackage{times}
\usepackage{booktabs}
\usepackage[letterpaper]{geometry}
\usepackage{url}
\usepackage{float}
\usepackage{graphicx}
\usepackage{setspace}
\usepackage{amsmath}
\usepackage{multirow}
\usepackage{algorithm}
\usepackage{algorithmic}

\newcommand{\eeqn}{\end{eqnarray*}}
\newcommand{\bneqn}{\vspace{-0.25cm}\begin{eqnarray}}
\newcommand{\eneqn}{\vspace{-0.25cm}\end{eqnarray}}

\providecommand{\tabularnewline}{\\}

\usepackage{pdflscape}
\author{Philip A. Ernst\footnote{Assistant Professor of Statistics, Department of Statistics, Rice University}\,, James R. Thompson\footnote{Noah Harding Emeritus Professor of Statistics, Department of Statistics, Rice University}\,, and Yinsen Miao\footnote{Ph.D. Candidate, Department of Statistics, Rice University}} 
\title{Portfolio selection: the power of equal weight}
\begin{document}
\maketitle

\begin{abstract}
We empirically show the superiority of the equally weighted S\&P 500 portfolio over Sharpe's market capitalization weighted S\&P 500 portfolio. We proceed to consider the MaxMedian rule, a non-proprietary rule designed for the investor who wishes to do his/her own investing on a laptop with the purchase of only 20 stocks. Rather surprisingly, over the 1958-2016 horizon, the cumulative returns of MaxMedian beat those of the equally weighted S\&P 500 portfolio by a factor of 1.15.
\end{abstract}

\section{Introduction}
The late John Tukey is well known for the following maxim (\cite{Tukey}):  ``far better an approximate answer to the right  question, which is often vague, than an exact answer to the wrong question, which can always be made precise.''

Let us consider the capital market theory of William Sharpe (\cite{Sharpe}) which assumes the following four axioms:
\begin{itemize}
\item The mean and standard deviation of a portfolio are sufficient for the purpose of investor
decision making.  
\item Investors can borrow and lend as much as they want at the risk-free rate of interest. 
\item All investors have the same expectations regarding the future, the same portfolios
available to them, and the same time horizon.
\item Taxes, transactions costs, inflation, and changes in interest rates may be ignored.
\end{itemize}

\begin{figure}[H]
\noindent \protect\centering{}\protect\includegraphics[width=8cm]{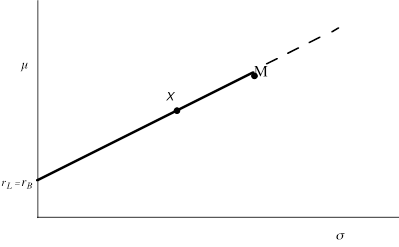}\protect\caption{The capital market line is given. The $y$-axis is the mean growth of the portfolio and the $x$-axis is the standard deviation of the portfolio.}
\protect
\label{fig1}
\end{figure}

\indent Based on these axioms, we present Figure \ref{fig1}. The mean growth of the portfolio is displayed on the $y$-axis and the standard deviation of the portfolio is displayed on the $x$-axis. According to capital market theory, if we start at the risk-free (T-bill) point $r_L$ and draw a straight line through the point $M$ (the market cap weighted portfolio
of all stocks), we should be unable to find any portfolio which lies above the capital market line.

\indent Each of the four axioms, however, is at best only an approximation to reality.  As noted by Tukey, frequently  the problem is not with the proof of a  theorem, but on the axioms assumed and their relationship to reality. Checking the conformity of the axioms to the real world is usually difficult. Rather, a better way to test the validity of a theorem is to test its effectiveness based on data.\\
\indent The authors of \cite{Thomp1} worked with real market data from 1970 through 2006. Working with the largest 1000 market cap stocks, they created 50,000 random funds for each year and compared their growth with the capital market line. They found that 66\% of the randomly generated funds lay above Sharpe's capital market line. In other words, that which Sharpe's axioms said could not happen did indeed happen in 66\% of the years. Figure \ref{fig2} below displays the data from one of these years (1993). The scatter plot displays funds for 1993. The capital market line is shown in green. According to Sharpe's axioms, all of the dots should fall below the green line,  but this is clearly not the case.
\begin{figure}[H]
\noindent \protect\centering{}\protect\includegraphics[width=8cm]{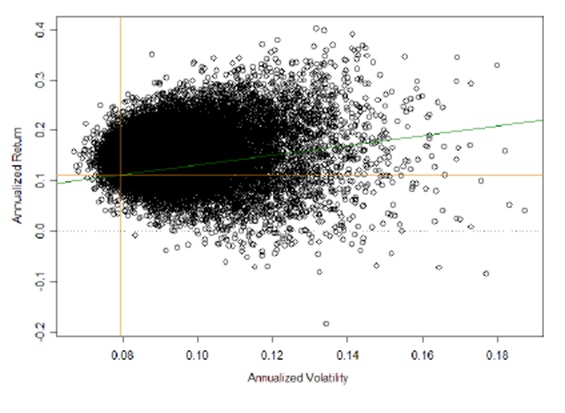}\protect\caption{Annualized return versus annualized volatility for 1993. Each dot represents a fund. The capital market line is in green.}
\protect
\label{fig2}
\end{figure}

In the foreword comments to \cite{Bogle} (an articulate defense of the S\&P 500 market capitalization weighted index fund),  Paul Samuelson writes ``Bogle's reasoned precepts can enable a few million of us savers to become in twenty years the envy of our suburban neighbors---while at the same time we have slept well in
these eventful times.'' To use a strategy which is beaten more often that not by a rather chaotic random strategy might not necessarily be a good thing (see \cite{thomp4}, among other sources). And yet, the S\&P 500 market cap weighted portfolio is probably used more than any other.\\
\indent Now it is not unusual in the statistical sciences to replace randomness by equal weight when a law is found not to be true. In the spirit of Tukey, we are not presently seeking
optimal weighting, but rather weighting which is superior to the market cap weighted
allocation in a portfolio. We will show, empirically, that both the equal weight and MaxMedian portfolios are generally superior to the market cap weighted portfolio.

\section{Data acquisition and index methodology} \label{sec2}
\subsection{Data acquisition}
This dataset considered in this paper is the S\&P 500 in the time frame from January 1958 to December 2016. All data were acquired from the Center for Research in Security Prices (CRSP)\footnote{http://www.crsp.com}. CRSP provides a reliable list of S\&P 500 index constituents, their respective daily stock prices, shares outstanding, and any key events (i.e., a stock split, an acquisition, etc.). For each constituent of the S\&P 500, CRSP also provides value weighted returns (both with and without dividends), equally weighted returns (both with and without dividends), index levels, and total market values. We utilize the Wharton Research Data Services (WRDS) interface\footnote{https://wrds-web.wharton.upenn.edu/wrds/support/index.cfm} to extract CRSP's S\&P 500 database into the R statistical programming platform. All plots and figures were produced using R.

\subsection{Index methodology}

The index return is the change in value of a portfolio over a given holding period.  We first calculate the index returns for both a equally weighted S\&P 500 portfolio and a market capitalization weighted S\&P 500 portfolio according to the index return formula as documented by CRSP \footnote{http://www.crsp.com/products/documentation/crsp-calculations}. CRSP computes the return on an index ($R_{t}$) as the weighted
average of the returns for the individual securities in the index according to the following equation
\begin{equation}
R_{t}=\dfrac{\sum_{i}\omega_{i,t}\times r_{i,t}}{\sum_{i}\omega_{i,t}}\label{eq:1},
\end{equation}
where $R_{t}$ is the index return, $\omega_{i,t}$ is the weight
of security $i$ at time $t$, and $r_{i,t}$ is the return of security
$i$ at time $t$.\\ 

\indent In a value-weighted index such as a market capitalization index (MKC), the weight $w_{i,t}$ assigned is its total market value, while in an equally weighted index (EQU), $w_{i,t}$ is set to one for each stock. Note that the security return $r_{i,t}$ can either be total return or capital appreciation (return without dividends). Whether it is the former or the latter determines, respectively, whether the index is a total return index or a capital appreciation index. In this paper we only consider the total return index.

\section{Equal weight and market cap portfolios} \label{sec3}

\subsection{Cumulative return of S\&P 500 from 1958 to 2016}
Using the CRSP database and the methodology documented in Section \ref{sec2}, we consider the cumulative returns for the S\&P 500 of the equally weighted S\&P 500 portfolio (EQU) and the market capitalization weighted S\&P 500 portfolio (MKC). Our calculation assumes that we invest \$100,000 (in 1958 dollars) in each of the EQU and MKC portfolios starting on 01/02/1958, and that the money is left to grow until 12/30/16. According to the Consumer Price Index (CPI) from Federal Reserve Bank of St. Louis\footnote{https://fred.stlouisfed.org/series/CPIAUCNS}, \$100,000 in 1958 dollars is approximately \$828,377.6 in 2016 dollars. \\
\indent Both portfolios are rebalanced monthly and the transaction fees are subtracted from the portfolio total at market close on the first trading day of every month. We assume transaction administrative fees of \$1 (in 2016 dollars) per trade and, additionally, a long-run average bid-ask spread of .1\% of the closing value of the stock. For example, if our portfolio buys (or sells) 50 shares of a given stock closing at \$100, transaction fees of $\$1+50\times(.1/2)=\$3.5$ is incurred. Dividend payments are \textit{included in the calculations}, both here and throughout the paper. \\
\indent Our main result appears in Figure \ref{fig3} below. Table \ref{table1} shows  that on 12/30/16 EQU is worth approximately \$172.89 million and that MKC is worth approximately \$38.44 million. Thus, EQU outperforms MKC in this time frame by a factor of 4.50.\\

\begin{figure}[H]
\noindent \centering
\includegraphics[clip,width=16cm]{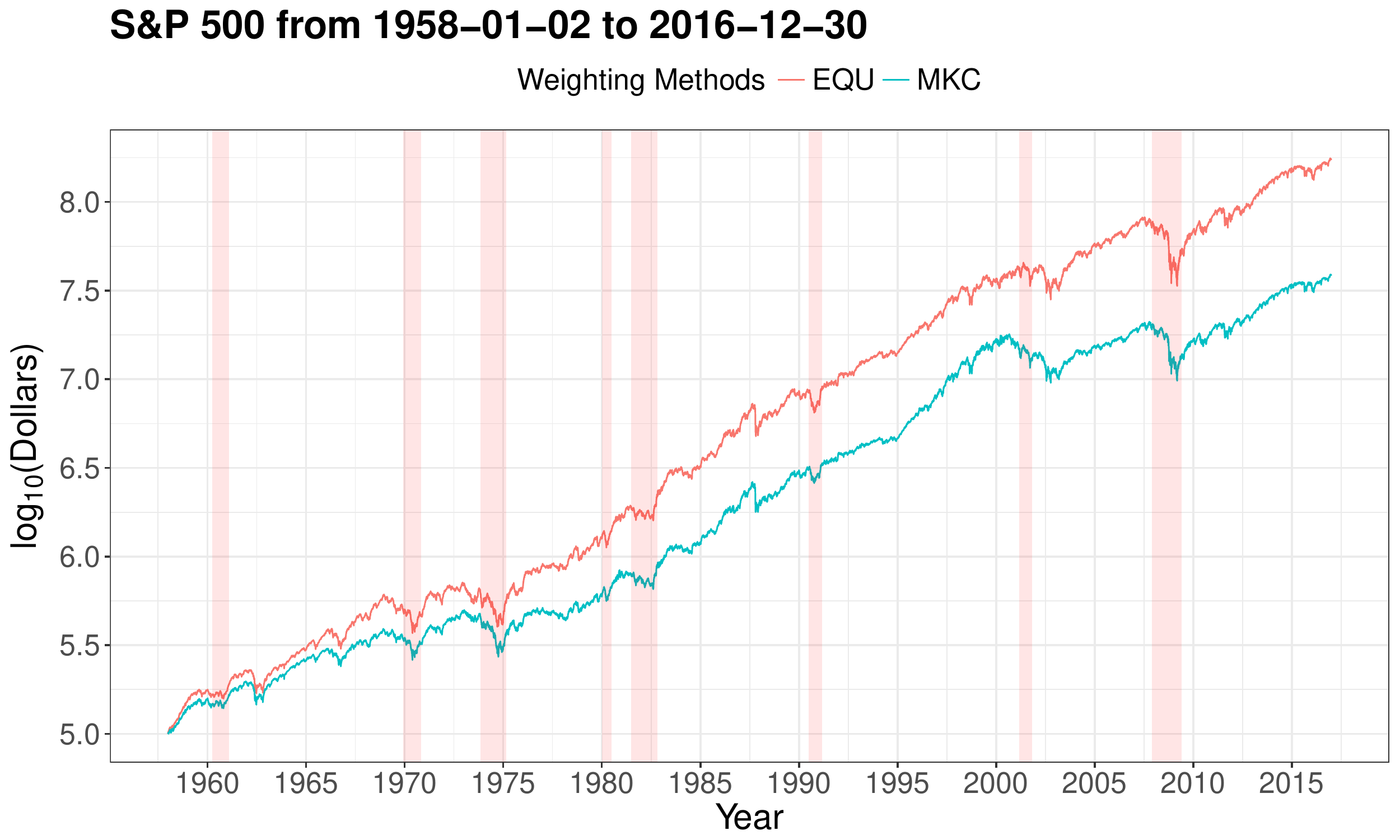}\caption{Cumulative return of S\&P 500 EQU and MKC portfolios.}
\label{fig3}
\end{figure}

\begin{table}[H]
\begin{onehalfspace}
\noindent \centering{}%
\begin{tabular}{llr}
\hline 
Date & EQU  & MKC \tabularnewline
\hline
2016-12-30  & \$172.89 mil & \$38.44 mil \tabularnewline
\hline 
\end{tabular}
\end{onehalfspace}
\caption{Values of S\&P 500 EQU and MKC portfolios on 12/30/2016.}
\label{table1}
\end{table}

\subsection{Transaction fees}
We provide a table of transaction fees incurred by EQU, MKC and MaxMedian (to be introduced in Section \ref{sec5}) over the 1958 to 2016 horizon. All numbers are discounted according to the CPI index. The total transaction fees are lowest for MaxMedian and largest for MKC. This is because MKC requires the most frequent rebalancing; MaxMedian, being that it is rebalanced annually, requires the least.

\begin{table}[H]
\begin{onehalfspace}
\begin{centering}
\begin{tabular}{r|r|r|r}
\hline 
 & EQU & MKC & Max-Median\tabularnewline
\hline 
Administration & \$.174 mil & \$.174 mil & \$4,810.44\tabularnewline
\hline 
Bid-ask Spread &\$.043 mil & \$.137 mil & \$48,104.43\tabularnewline
\hline 
Total & \$.217 mil & \$.311 mil & \$52,914.87\tabularnewline
\hline 
\end{tabular}\caption{Administration fee (\$1 per trade) and bid-ask spread (0.1\% of the
closing price per stock) for each of the three portfolios under consideration
from 1958-2016. EQU and MKC are rebalanced monthly. The MaxMedian portfolio
is rebalanced annually.}
\par\end{centering}
\end{onehalfspace}
\end{table}

\begin{figure}[H]
\noindent \centering
\includegraphics[clip,width=14cm]{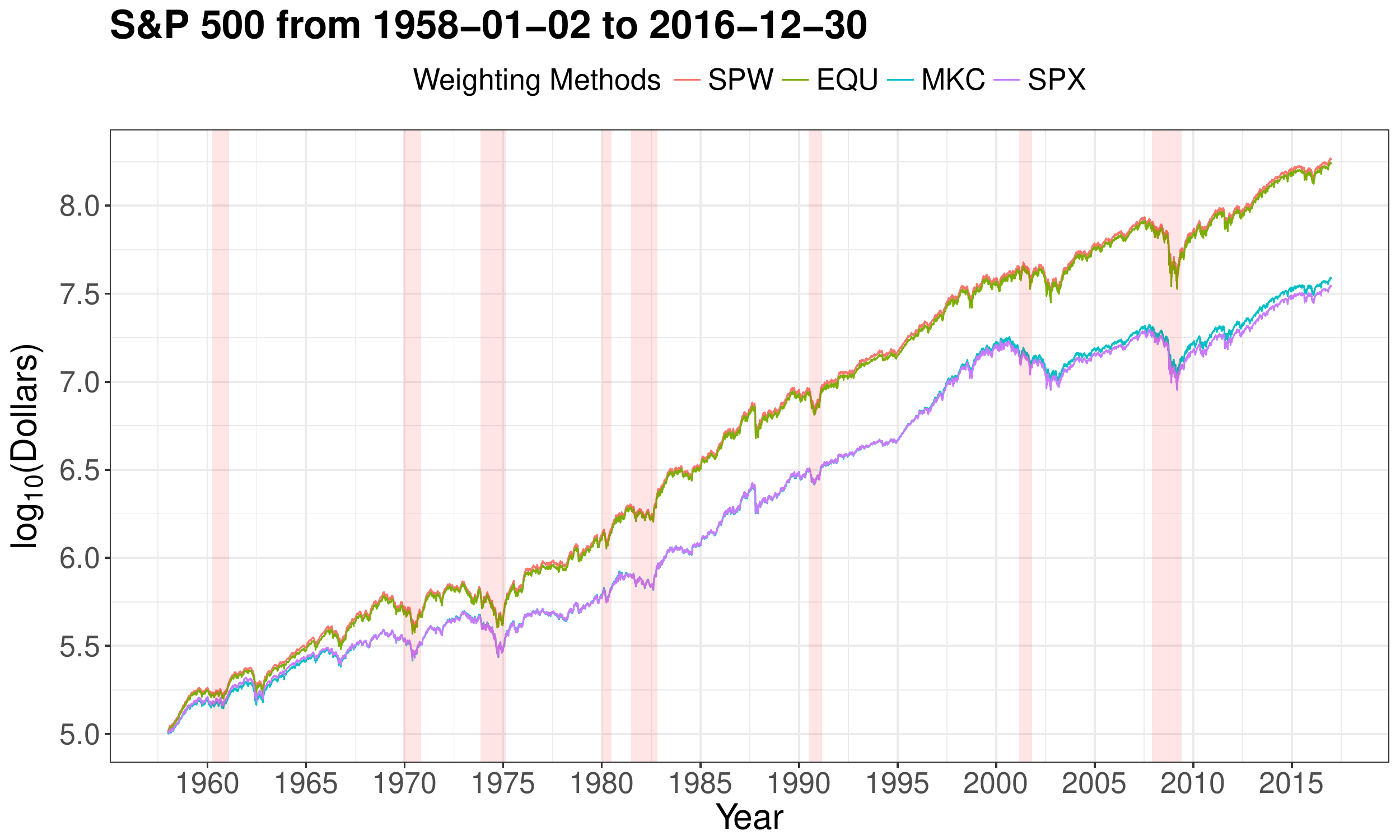}\caption{Comparison of cumulative returns between MKC and EQU with, respectively, SPX and SPW (CRSP's equivalent portfolios).}
\label{fig3_1}
\end{figure}

The above results can be rigorously cross-validated. From January 1926 to the present, CRSP has calculated daily returns of their defined S\&P 500 market capitalization portfolio (SPX) as well as their defined S\&P 500 equally weighted portfolio (SPW). Using CRSP's reported daily returns with dividends, we find that our cumulative EQU figure (\$172.89 million) is close to the figure obtained by SPW (\$182.19 million). Similarly, our cumulative MKC figure (\$38.44 million) is close to that of SPX (\$34.69 million).  The differences may attributed to the different methods for computing the transaction and bid-ask spread fees. We assume the transaction administrative fees of \$1 (in 2016 dollars) per trade and a long-run average bid-ask spread of .1\% of the closing price of the stock. CRSP's methods, however, are not public.  A comparison plot of cumulative return between the MKC and EQU portfolios and, respectively, CRSP's SPX and SPW is given in Figure \ref{fig3_1}. As indicted in Figure \ref{fig3_1}, the cumulative returns of MKC and SPX as well as those of EQU and SPW are indeed very close.
\vspace{0.5cm}

\subsection{Annual rates of return for EQU and MKC}
In Table \ref{table2} below, we display annual returns (in percentage) of the EQU and MKC portfolios. We also display the annual returns of SPW (CRSP's equal weight portfolio) and SPX (CRSP's market cap weight portfolio). The returns of EQU and MKC are in close agreement with the reported returns of SPW and SPX. Returning to our discussion of Table \ref{table2}, we see below that the geometric mean of EQU is 13.47\% and the geometric mean of MKC is  10.61\%, giving EQU a substantial annual advantage of  2.86\%. In addition, Sharpe ratios are computed using a risk-free rate of 1.75\%. The Sharpe ratio for EQU beats that of MKC by a factor of 1.16.

\begin{table}[H]
\begin{onehalfspace}
\begin{centering}
\begin{tabular}{cr@{\extracolsep{0pt}.}lr@{\extracolsep{0pt}.}lr@{\extracolsep{0pt}.}lr@{\extracolsep{0pt}.}lcr@{\extracolsep{0pt}.}lr@{\extracolsep{0pt}.}lr@{\extracolsep{0pt}.}lr@{\extracolsep{0pt}.}l}
\toprule
\hline
Year & \multicolumn{2}{c}{EQU} & \multicolumn{2}{c}{SPW } & \multicolumn{2}{c}{MKC} & \multicolumn{2}{c}{SPX} & Year & \multicolumn{2}{c}{EQU} & \multicolumn{2}{c}{SPW } & \multicolumn{2}{c}{MKC} & \multicolumn{2}{c}{SPX}\tabularnewline
\hline
\midrule
1958  & 54&84  & 55&47  & 41&66  & 42&40  & 1987  & 7&95  & 8&29  & 5&62  & 5&05 \tabularnewline
1959  & 13&90  & 14&40  & 11&51  & 12&49  & 1988  & 22&05  & 22&33  & 16&72  & 17&02 \tabularnewline
1960  & -0&73  & -0&38  & -1&72  & 0&57  & 1989  & 26&96  & 27&15  & 31&22  & 31&40 \tabularnewline
1961  & 29&39  & 29&65  & 25&96  & 27&20  & 1990  & -11&22  & -11&25  & -2&76  & -3&24 \tabularnewline
1962  & -10&80  & -10&44  & -7&82  & -8&78  & 1991  & 37&12  & 37&14  & 30&38  & 30&69 \tabularnewline
1963  & 23&62  & 24&06  & 22&28  & 22&69  & 1992  & 15&63  & 15&66  & 7&81  & 7&68 \tabularnewline
1964  & 19&54  & 19&92  & 17&99  & 16&69  & 1993  & 15&70  & 15&70  & 10&24  & 9&78 \tabularnewline
1965  & 24&49  & 24&81  & 14&19  & 12&60  & 1994  & 1&63  & 1&62  & 1&56  & 1&39 \tabularnewline
1966  & -8&02  & -8&26  & -9&81  & -10&25  & 1995  & 32&83  & 32&86  & 38&06  & 37&59 \tabularnewline
1967  & 37&09  & 37&17  & 26&31  & 24&01  & 1996  & 20&43  & 20&44  & 24&84  & 23&19 \tabularnewline
1968  & 26&60  & 26&64  & 11&19  & 11&00  & 1997  & 29&35  & 29&34  & 34&45  & 33&45 \tabularnewline
1969  & -17&47  & -17&22  & -8&46  & -8&23  & 1998  & 13&96  & 13&86  & 29&44  & 29&03 \tabularnewline
1970  & 7&12  & 6&54  & 3&82  & 4&22  & 1999  & 12&36  & 12&30  & 22&11  & 20&94 \tabularnewline
1971  & 19&00  & 17&98  & 15&37  & 14&04  & 2000  & 10&91  & 10&85  & -7&31  & -8&99 \tabularnewline
1972  & 11&56  & 11&12  & 19&13  & 19&22  & 2001  & 1&72  & 1&69  & -11&84  & -11&80 \tabularnewline
1973  & -21&23  & -21&32  & -14&43  & -15&12  & 2002  & -16&44  & -16&56  & -21&24  & -22&10 \tabularnewline
1974  & -20&92  & -21&13  & -27&50  & -26&47  & 2003  & 42&18  & 42&20  & 28&59  & 28&71 \tabularnewline
1975  & 54&24  & 57&76  & 37&27  & 36&52  & 2004  & 17&56  & 17&56  & 10&83  & 10&95 \tabularnewline
1976  & 36&25  & 36&71  & 23&10  & 23&95  & 2005  & 7&93  & 7&96  & 5&09  & 5&10 \tabularnewline
1977  & -1&19  & -1&35  & -7&87  & -7&44  & 2006  & 16&37  & 16&33  & 15&73  & 15&65 \tabularnewline
1978  & 9&97  & 9&27  & 6&90  & 6&29  & 2007  & 0&86  & 0&87  & 5&58  & 5&74 \tabularnewline
1979  & 30&16  & 29&71  & 19&94  & 18&55  & 2008  & -38&09  & -37&99  & -35&01  & -36&65 \tabularnewline
1980  & 31&57  & 31&76  & 33&45  & 32&62  & 2009  & 48&93  & 48&97  & 27&58  & 26&21 \tabularnewline
1981  & 6&09  & 5&23  & -6&76  & -5&08  & 2010  & 22&28  & 22&26  & 15&47  & 15&12 \tabularnewline
1982  & 31&13  & 31&70  & 21&87  & 21&96  & 2011  & 0&24  & 0&26  & 1&82  & 1&84 \tabularnewline
1983  & 31&77  & 31&47  & 22&72  & 22&33  & 2012  & 17&56  & 17&57  & 16&10  & 16&11 \tabularnewline
1984  & 3&90  & 4&13  & 5&81  & 6&67  & 2013  & 36&33  & 36&39  & 32&22  & 32&36 \tabularnewline
1985  & 31&69  & 31&97  & 31&99  & 32&04  & 2014  & 14&46  & 14&44  & 13&70  & 13&59 \tabularnewline
1986  & 18&68  & 19&03  & 17&96  & 18&30  & 2015  & -2&30  & -2&23  & 1&47  & 1&49 \tabularnewline
 & \multicolumn{2}{c}{} & \multicolumn{2}{c}{} & \multicolumn{2}{c}{} & \multicolumn{2}{c}{} & 2016  & 15&42  & 15&56  & 11&79  & 11&81 \tabularnewline

\midrule
\hline
 & \multicolumn{2}{c}{} & \multicolumn{2}{c}{} & \multicolumn{2}{c}{} & \multicolumn{3}{r}{\textbf{Arithmetic}} & 15&13 & 15&22 & 11&97 & 11&77\tabularnewline
 & \multicolumn{2}{c}{} & \multicolumn{2}{c}{} & \multicolumn{2}{c}{} & \multicolumn{3}{r}{\textbf{Geometric}} & 13&47 & 13&53 & 10&61 & 10&40\tabularnewline
 & \multicolumn{2}{c}{} & \multicolumn{2}{c}{} & \multicolumn{2}{c}{} & \multicolumn{3}{r}{\textbf{SD}} & 19&01 & 19&20 & 16&83 & 16&80\tabularnewline
 & \multicolumn{2}{c}{} & \multicolumn{2}{c}{} & \multicolumn{2}{c}{} & \multicolumn{3}{r}{\textbf{Sharpe Ratio}} & 70&38 & 70&16 & 60&72 & 59&64\tabularnewline
\bottomrule
\hline
\end{tabular}
\par\end{centering}
\end{onehalfspace}
\caption{Annual rates of return (in \%) for the EQU, SPW, MKC and SPX portfolios. The arithmetic and geometric means, standard deviation,
and Sharpe ratios (in \%) of the annual returns for all four portfolios (over
the full 1958-2016 time horizon) are also listed.}
\label{table2} 
\end{table}

\newpage

\section{The ``everyday'' investor: a simple rule} \label{sec4}
Despite the low expense ratios of many mutual funds, many personal investors prefer not to invest in a mutual fund. A personal investor may choose to invest in twenty stocks.  Such an investor could choose to invest in the top 20 stocks by market capitalization of the S\&P 500, the middle 20 stocks (stocks 241-260) by market capitalization, or the bottom 20 stocks by market capitalization. What would happen to an investor who chooses one of these three aforementioned 20 stock baskets and then invests in these according to equal weight or according to market capitalization weight?\\

\indent As in Section \ref{sec3}, we continue to assume that \$100,000 is invested in each portfolio on 1/2/1958, and that it is left to grow in the portfolio until 12/30/2016. CRSP's method given in equation \mbox{(\ref{eq:1})} is used to compute the cumulative returns. The cumulative returns for the top 20 stock basket, the middle 20 stock basket, and the bottom 20 stock basket, using both EQU and MKC portfolio weighting, is given below. 

\begin{table}[H]
\begin{doublespace}
\noindent \begin{centering}
\begin{tabular}{c|r@{\extracolsep{0pt}.}l|r@{\extracolsep{0pt}.}l|r@{\extracolsep{0pt}.}l|r@{\extracolsep{0pt}.}l}
\hline 
 & \multicolumn{2}{c|}{Top 20} & \multicolumn{2}{c|}{Middle 20} & \multicolumn{2}{c|}{Bottom 20} & \multicolumn{2}{c}{Full Data}\tabularnewline
\hline 
EQU & \$150&93 mil & \$21&97 mil & \$0&29 mil & \$172&89 mil\tabularnewline
\hline 
MKC & \$51&21 mil & \$5&55 mil & \$2748&44& \$38&44 mil\tabularnewline
\hline 
\end{tabular}
\par\end{centering}
\end{doublespace}

\caption{Top, middle, and bottom 20 baskets of EQU and MKC and their cumulative returns
from January 1958 to December 2016.}
\label{table3}
\end{table}

From Table \ref{table3}, we may conclude as follows:
\begin{enumerate}
\item For each of the top 20, mid 20, and bottom 20 stock baskets, equal weighting significantly trumps market capitalization weighting. It does so by a factor of 2.95 for the top 20 stock basket, by a factor of 3.92 for the mid 20 stock basket, and by a factor of about 107.20 for the bottom 20 stock basket. 
\item Investing equally in all 500 stocks of the S\&P 500 outperforms the top 20 equal weight stock picker by a factor of 1.14 for the 1958-2016 time range. However, investing according to market capitalization in all 500 stocks of the S\&P 500 underperforms the top 20 market capitalization weight stock picker (the latter outperforms by a factor of 1.33 for the 1958-2016 time range).
\end{enumerate}

\section{MaxMedian: achieving returns superior to EQU using only a laptop} \label{sec5}
The MaxMedian rule is a non-proprietary rule which was designed for the investor who wishes to do his/her own investing on a laptop with the purchase of only 20 stocks.  The rule, which was first discovered by the second author in \cite{thomp3} and was further documented in \cite{thomp2}, is summarized below:

\begin{enumerate}
\item For the 500 stocks in the S\&P 500, obtain the daily prices $S(j,t)$ for the preceding year.
\item Compute daily ratios as follows: $r(j,t) = S(j,t)/S(j,t$-$1).$
\item Sort these ratios for the year's trading days. 
\item Discard all values of $r$ equal to one.
\item Examine the 500 medians of the ratios.
\item Invest equally in the 20 stocks with the largest medians.
\item Hold for one year and then liquidate. Repeat steps 1-6 again for each future year.
\end{enumerate}

According to the second author, MaxMedian was not implemented with any hope that it would beat equal weight on the S\&P 500.  However, as we see in Figure \ref{fig5} below, this is indeed the case. Figure \ref{fig5} below shows the cumulative return for EQU, MKC, and the MaxMedian rule. As before, dividends and transaction fees are included in all calculations.

\begin{figure}[H]
\noindent \centering{}\includegraphics[clip,width=14cm]{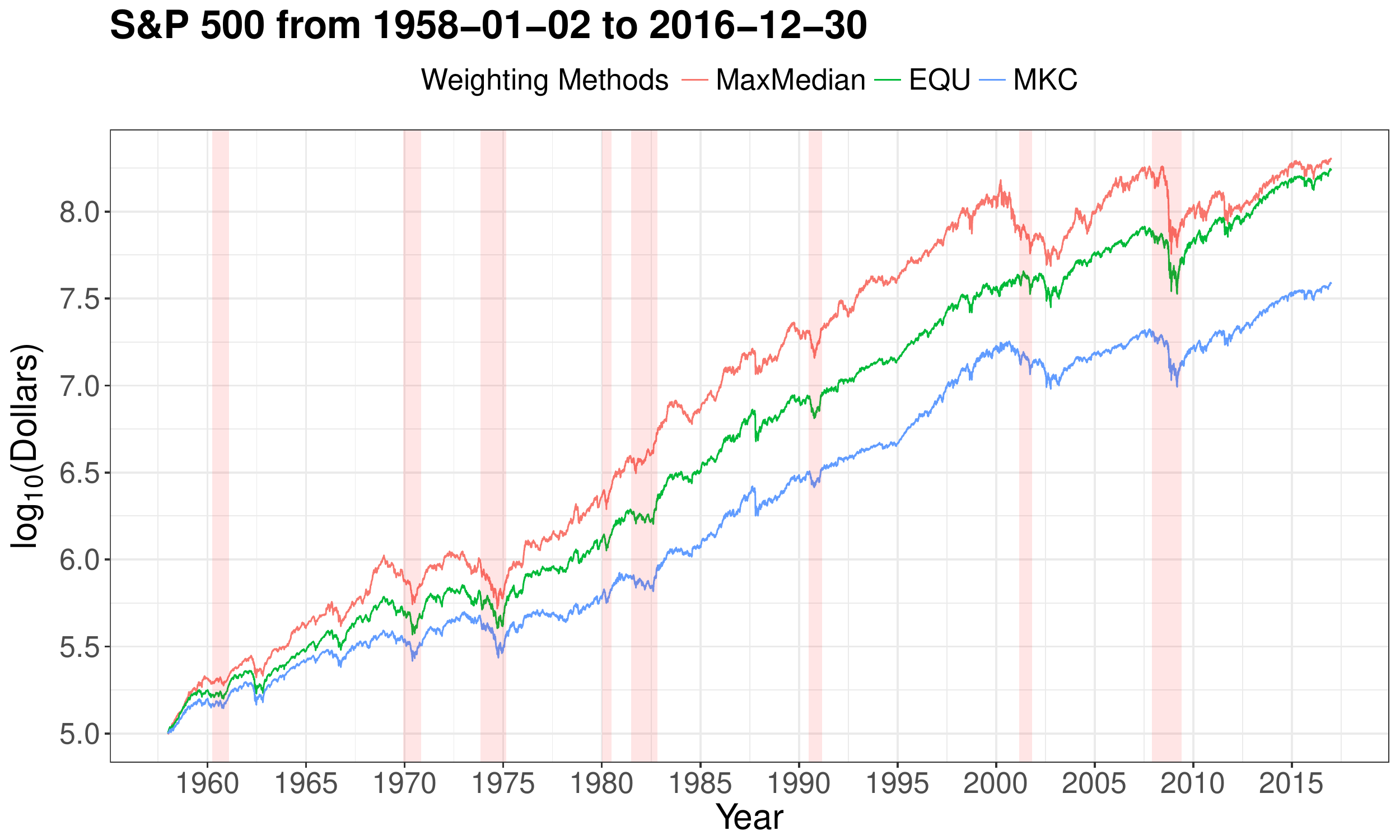}\caption{Cumulative return of S\&P 500 for EQU, MKC, and MaxMedian}
\label{fig5}
\end{figure}

\begin{table}[H]
\begin{onehalfspace}
\noindent \centering{}%
\begin{tabular}{rrrr@{\extracolsep{0pt}.}l}
\hline 
\multirow{1}{*}{Date } & \multirow{1}{*}{EQU } & \multirow{1}{*}{MKC } & \multicolumn{2}{c}{\multirow{1}{*}{MaxMedian }}\tabularnewline
\hline 
2016-12-30  & \$172.89 mil & \$38.44 mil & \$199.41mil\tabularnewline
\hline 
\end{tabular}\caption{Return of S\&P 500 for EQU, MKC, and MaxMedian on 12/30/2016.}
\label{table4}
\end{onehalfspace}
\end{table}

Table \ref{table4} shows that the cumulative return for MaxMedian (\$199.41mil) beats EQU (\$172.89 mil) by a factor of 1.15. In Table \ref{table5}, we provide the annual returns for EQU, MKC, and MaxMedian.  Sharpe ratios are computed using a risk-free rate of 1.75\%. It should be noted Sharpe ratio for EQU is superior to that of MaxMedian, as the latter has a higher annual standard deviation.
\begin{table}[H]
\begin{onehalfspace}
\begin{centering}
\begin{tabular}{cr@{\extracolsep{0pt}.}lr@{\extracolsep{0pt}.}lr@{\extracolsep{0pt}.}lcr@{\extracolsep{0pt}.}lr@{\extracolsep{0pt}.}lr@{\extracolsep{0pt}.}l}
\hline
\toprule 
\textbf{Year} & \multicolumn{2}{c}{\textbf{EQU}} & \multicolumn{2}{c}{\textbf{MKC}} & \multicolumn{2}{c}{\textbf{MaxMedian}} & \textbf{Year} & \multicolumn{2}{c}{\textbf{EQU}} & \multicolumn{2}{c}{\textbf{MKC}} & \multicolumn{2}{c}{\textbf{MaxMedian}}\tabularnewline
\hline
\midrule
1958  & 54&84  & 41&66  & 62&46  & 1987  & 7&95  & 5&62  & 1&62 \tabularnewline
1959  & 13&90  & 11&51  & 27&80  & 1988  & 22&05  & 16&72  & 23&87 \tabularnewline
1960  & -0&73  & -1&72  & -2&39  & 1989  & 26&96  & 31&22  & 32&84 \tabularnewline
1961  & 29&39  & 25&96  & 29&69  & 1990  & -11&22  & -2&76  & -12&81 \tabularnewline
1962  & -10&80  & -7&82  & -3&32  & 1991  & 37&12  & 30&38  & 56&19 \tabularnewline
1963  & 23&62  & 22&28  & 24&98  & 1992  & 15&63  & 7&81  & 16&87 \tabularnewline
1964  & 19&54  & 17&99  & 29&01  & 1993  & 15&70  & 10&24  & 22&64 \tabularnewline
1965  & 24&49  & 14&19  & 18&75  & 1994  & 1&63  & 1&56  & 1&44 \tabularnewline
1966  & -8&02  & -9&81  & -4&67  & 1995  & 32&83  & 38&06  & 26&28 \tabularnewline
1967  & 37&09  & 26&31  & 34&59  & 1996  & 20&43  & 24&84  & 25&31 \tabularnewline
1968  & 26&60  & 11&19  & 60&73  & 1997  & 29&35  & 34&45  & 28&82 \tabularnewline
1969  & -17&47  & -8&46  & -23&60  & 1998  & 13&96  & 29&44  & 24&77 \tabularnewline
1970  & 7&12  & 3&82  & 3&08  & 1999  & 12&36  & 22&11  & 16&71 \tabularnewline
1971  & 19&00  & 15&37  & 27&48  & 2000  & 10&91  & -7&31  & -33&83 \tabularnewline
1972  & 11&56  & 19&13  & 4&70  & 2001  & 1&72  & -11&84  & -8&86 \tabularnewline
1973  & -21&23  & -14&43  & -18&66  & 2002  & -16&44  & -21&24  & -18&92 \tabularnewline
1974  & -20&92  & -27&50  & -27&24  & 2003  & 42&18  & 28&59  & 49&82 \tabularnewline
1975  & 54&24  & 37&27  & 52&55  & 2004  & 17&56  & 10&83  & 12&02 \tabularnewline
1976  & 36&25  & 23&10  & 38&97  & 2005  & 7&93  & 5&09  & 39&93 \tabularnewline
1977  & -1&19  & -7&87  & 9&79  & 2006  & 16&37  & 15&73  & 7&60 \tabularnewline
1978  & 9&97  & 6&90  & 19&06  & 2007  & 0&86  & 5&58  & 7&49 \tabularnewline
1979  & 30&16  & 19&94  & 33&33  & 2008  & -38&09  & -35&01  & -51&28 \tabularnewline
1980  & 31&57  & 33&45  & 39&30  & 2009  & 48&93  & 27&58  & 32&83 \tabularnewline
1981  & 6&09  & -6&76  & 16&56  & 2010  & 22&28  & 15&47  & 17&84 \tabularnewline
1982  & 31&13  & 21&87  & 53&40  & 2011  & 0&24  & 1&82  & -17&78 \tabularnewline
1983  & 31&77  & 22&72  & 30&11  & 2012  & 17&56  & 16&10  & 13&64 \tabularnewline
1984  & 3&90  & 5&81  & -1&66  & 2013  & 36&33  & 32&22  & 35&27 \tabularnewline
1985  & 31&69  & 31&99  & 34&12  & 2014  & 14&46  & 13&70  & 20&30 \tabularnewline
1986  & 18&68  & 17&96  & 26&44  & 2015  & -2&30  & 1&47  & -3&52 \tabularnewline
 & \multicolumn{2}{c}{} & \multicolumn{2}{c}{} & \multicolumn{2}{c}{} & 2016  & 15&42  & 11&79  & 9&81 \tabularnewline
 \hline
\midrule
 & \multicolumn{2}{c}{} & \multicolumn{2}{c}{} & \multicolumn{3}{r}{\textbf{Arithmetic}} & 15&13 & 11&97 & 16&48\tabularnewline
 & \multicolumn{2}{c}{} & \multicolumn{2}{c}{} & \multicolumn{3}{r}{\textbf{Geometric}} & 13&47 & 10&61 & 13&75\tabularnewline
 & \multicolumn{2}{c}{} & \multicolumn{2}{c}{} & \multicolumn{3}{r}{\textbf{SD}} & 19&01 & 16&83 & 23&87\tabularnewline
 & \multicolumn{2}{c}{} & \multicolumn{2}{c}{} & \multicolumn{3}{r}{\textbf{Sharpe Ratio}} & 70&38 & 60&72 & 61&71\tabularnewline
 \hline
\end{tabular}
\par\end{centering}
\end{onehalfspace}
\caption{Annual rates of return (in \%) for the EQU, MKC and MaxMedian portfolios.
The arithmetic and geometric means, standard deviation, and Sharpe
ratios (in \%) of annual return all three portfolios (over the full
1958-2016 time period) are listed as well.}
\label{table5} 
\end{table}


Returning to the geometric mean, note that the geometric mean of EQU is 13.47\% and the geometric mean of MaxMedian is  13.75\%, giving MaxMedian an annual advantage over EQU of  .28\%. One can continue searching for weighting schemes which do even better than MaxMedian, which we do in \cite{Ernst}. We do however again stress the importance of achieving this performance with a portfolio size of only 20 securities.

\section{Supplementary Materials}
For purposes of replicability, all data used in this work can be found online on the following GitHub repository: \url{https://github.com/yinsenm/equalitySP500}.


\bibliographystyle{plain}
\bibliography{VF}
\end{document}